\documentclass[%
 reprint,
nofootinbib,
 amsmath,amssymb,
 aps,
]{revtex4-2}
\usepackage[utf8]{inputenc}
\usepackage{float}

\usepackage{graphicx, amsmath,amsfonts,amssymb}
\usepackage{braket}

\usepackage{sidecap}
\usepackage{comment}
\usepackage{hyperref}% add hypertext capabilities
\hypersetup{
    colorlinks=true,
    linkcolor=blue,
    urlcolor=blue,
    citecolor=red
}

\begin{document}

\title{Entanglement transitions with free fermions}
\date{October 2022}
\author{Joseph Merritt}
\author{Lukasz Fidkowski}
\affiliation{Department of Physics, University of Washington, Seattle, WA 98195-1560, USA}

\begin{abstract}
    We use Majorana operators to study entanglement dynamics under random free fermion unitary evolution and projective measurements in one dimension.  For certain choices of unitary evolution, namely those which swap neighboring Majorana operators, and measurements of neighboring Majorana bilinears, one can map the evolution to the statistical model of completely packed loops with crossings (CPLC) and study the corresponding phase diagram. 
    %This phase diagram exhibits a transition from an area law to a critical Goldstone phase with a logarithmic entropy scaling law. 
    We generalize this model using the language of fermionic Gaussian states to a general free fermion unitary evolution acting on neighboring Majorana operators, and numerically compute its phase diagram.  We find that both the Goldstone and area law phases persist in this new phase diagram, but with a shifted phase boundary.  One important qualitative aspect of the new phase boundary is that even for the case of commuting measurements, the Goldstone phase persists up to a finite non-zero measurement rate.  This is in contrast with the CPLC, in which non-commuting measurements are necessary for realizing the Goldstone phase.  We also numerically compute the correlation length critical exponent at the transition, which we find to be near to that of the CPLC, and give a tentative symmetry based explanation for some differences in the phase transition line between the CPLC and generalized models.
    
\end{abstract}

\maketitle

\section{Introduction}

Recently the study of `hybrid' quantum circuits, involving both unitary dynamics and projective measurements, has received a great deal of attention \cite{Fisher,et2, etreview}.  By focusing on the ensemble of quantum trajectories of pure states defined by the various measurement outcomes one can study new types of non-equilibrium phase transitions, with the canonical example being the `entanglement transition'.  In the entanglement transition, the ensemble-averaged entanglement entropy changes from scaling as an area law to scaling as a volume law as the measurement rate is decreased.  A closely related concept is that of a `purification' transition, where instead of pure quantum state trajectories one studies the purifying behavior of an initial maximally mixed state.

In the special case of free fermion dynamics \cite{Jian} - i.e. unitaries which are exponentials of bilinears of the creation and annihilation operators, and measurements only of Fock space mode occupation numbers - the mixed (or `volume law') phase is known to be unstable to any non-zero measurement rate \cite{Fidkowski}.  However, such free fermion dynamics can still accommodate an interesting phase transition from a purifying phase to a so-called `Goldstone' phase \cite{Nahum}.  The latter still exhibits purifying behavior, but on time scales parametrically longer in system size.  More precisely, in the Goldstone phase the entropy for a system of length $L$ after a time of order $L$ scales like $\log L$ (whereas in the purifying phase this entropy would be close to $0$, i.e. the state would have approximately purified long before this time scale was reached).  This phase transition can be realized in a specific free fermion model, one that can be solved \cite{Nahum_Skinner} by exact mapping to a known statistical mechanical model, the completely packed loop model with crossings (CPLC) \cite{Nahum}.

A natural question one may ask is, how generic is the CPLC phase diagram in the context of free fermion hybrid dynamics?  In other words, do the area law and Goldstone phases persist when the dynamics is deformed slightly away from the specific point dual to the CPLC model?  Is the phase transition still continuous, and is it in the same universality class as that of the CPLC model?

In this work, we investigate these questions by extending the CPLC-dual free fermion model to a more general family of free fermion models.  Specifically, following \cite{Nahum_Skinner}, the CPLC-dual model has a convenient description in terms of a one dimensional chain of Majorana fermions, as follows: the unitary gates swap a neighboring pair of Majoranas ($\gamma_1 \rightarrow \gamma_2, \gamma_2\rightarrow -\gamma_1$), and the measurements measure the occupation number of a free fermion mode defined by a neighboring pair of Majoranas.  These gates are implemented with respect to one pairing of Majoranas and its complementary pairing in an alternating fashion, as described in detail below.  The phase diagram is a function of two parameters, $p$ and $q$, which control, respectively, the rate of measurements and the asymmetry between the two complementary pairings of Majoranas.  Our generalized model replaces the swap gate, which may be thought of as a $\frac{\pi}{2}$ rotation in the $SO(2)$ that rotates $\gamma_1$ into $\gamma_2$, by a rotation by a random angle inside this $SO(2)$.  The measurement gates are as in the CPLC model, and the phase diagram is once again a function of the two parameters $p$ and $q$. 

Our generalized model no longer admits an easily solvable statistical mechanical dual, although in principle some statistical mechanical dual should exist, as discussed below.  To study it, we therefore instead leverage the free fermion nature of the dynamics to perform efficient Monte Carlo simulations using the Gaussian state formalism.  The essential feature of both the CPLC and our generalized models which makes this possible is the fact that, for a particular quantum trajectory, the many-body quantum state of $2N$ Majoranas remains Gaussian, meaning that it can be efficiently encoded in a correlation matrix with $O(N^2)$ entries.  This allows us to avoid having to simulate dynamics in a Hilbert space exponentially large in $N$.  Ultimately this is just a consequence of the free fermion nature of the dynamics.

We find that the general features of the CPLC phase diagram persist in the generalized model.  The area law and Goldstone phases remain, but the phase transition between them shifts slightly.  An important qualitative difference is that the Goldstone phase persists down to finite $p<1$ for $q=0,1$ in the generalized model, in contrast to the CPLC.  This implies that the unitaries in the generalized model are more scrambling in some sense than those of the CPLC, because they can support the Goldstone phase with commuting projectors (i.e. at $q=0,1$).  The CPLC, on the other hand, requires non-commuting measurements \cite{Barkeshli, Hsieh, Lang, Khemani} to support the Goldstone phase.  Our result is consistent with the fact that a volume entanglement law cannot be maintained in free fermi systems at finite measurement rates \cite{DeLuca, Fidkowski}, and comports with the results found in \cite{Diehl} in the case of continuous monitoring, and bears resemblance to the results of \cite{Lucas} and \cite{Tang} in the case of non-unitary free fermion evolution (see also \cite{Kells} for additional exploration of free fermion phases and phase transitions with weak measurements, and \cite{Schiro1, Schiro2} for further study of the phase transition from an area law phase to a logarithmic phase).  We would also like to note that the transitions found in the CPLC have also been studied using entanglement negativity and other measures in the context of monitored dynamics in \cite{Sang}.

We also perform a finite size scaling analysis that allows us to extract a correlation length critical exponent $\nu\approx 2.4$ for the generic transition between the two phases.  The accuracy of our analysis is not sufficient to definitively conclude that this corresponds to a different universality class from the CPLC model, which has $\nu_{\text{CPLC}} \approx 2.75$ \cite{Nahum}.

The rest of this paper is structured as follows. In Section \ref{sec:model} we review the CPLC model and construct the duality mapping between this model and a free fermion hybrid dynamics.  In particular, we highlight the connection between the `spanning number' in the CPLC model and the entropy in the quantum model. In Section \ref{sec:gaussian} we discuss more general free fermion models, and introduce the Guassian state formalism that allows us to efficiently simulate them.  In Section \ref{sec:results} we present the results of our Monte Carlo numerical simulations of the more general free fermion models. In Section \ref{sec:discussion} we summarize our results and consider future directions.  In particular, we discuss a qualitative change in the shape of the phase boundary between the CPLC-dual and generalized models, and propose a symmetry-based explanation of this difference.

\section{Exactly solvable model of free fermion hybrid dynamics} \label{sec:model}

This Section outlines a particular implementation of the duality between the CPLC and a quantum model of Majorana worldlines, first proposed in \cite{Nahum_Skinner}.  We consider a one dimensional chain of $N$ spinless fermions, and write the operator algebra in terms of $2N$ Majorana fermions $\gamma_k$ ($k=1,\ldots,2N$).  These are related to the usual creation and annihilation operators $a_j, a_j^\dagger$ ($j=1,\ldots,N$) by:

\begin{align*}
    \gamma_{2j-1} &= a_j + a^\dagger_j \\
    \gamma_{2j} &= i(a_j - a^\dagger_j)
\end{align*}
\begin{align*}
    a_j &= \frac{1}{2} \left(\gamma_{2j-1} - i \gamma_{2j}\right) \\
    a_j^\dagger &= \frac{1}{2} \left(\gamma_{2j-1} + i \gamma_{2j} \right)
\end{align*}
We have $i\gamma_{2j-1} \gamma_{2j}=(-1)^{n_j} = 1-2n_j$ where $n_j = a^\dagger_j a_j$ is the occupation number operator at site $j$, taking eigenvalues $0$ and $1$.

\subsection*{Hybrid unitary-measurement circuit}

We take periodic boundary conditions, so that a subscript of $N+1$ below is to be interpreted as $1$.  The time step is labeled by a positive integer, and the protocol depends on the parity of this time step.  $p$ and $q$ are two real numbers between $0$ and $1$ which serve as control parameters.  For convenience, let us first define the two-Majorana unitary gate $U_{r,r+1}$, ($r=1,\ldots,2N$) by
\begin{align}
U_{r,r+1}=\frac{1}{\sqrt{2}} \left(1-\gamma_{r}\gamma_{r+1}\right)
\end{align}
This gate acts as follows:

\begin{align*}
U_{r,r+1}\gamma_r U_{r,r+1}^\dagger&=\gamma_{r+1}\\
U_{r,r+1}\gamma_{r+1} U_{r,r+1}^\dagger&=-\gamma_r
\end{align*}
while commuting with all $\gamma_j$, $j\neq r,r+1$.

{\bf Odd time steps:} We perform 2-Majorana gates on all pairs $(2j-1,2j)$ $(j=1,\ldots,N)$ of nearest neighbor Majoranas.  For each such pair $(2j-1,2j)$ the gate is chosen randomly from $3$ possibilities: (1) with probability $p$ we act with $U_{2j-1,2j}$; (2) with probability $(1-p)q$ we measure $i\gamma_{2j-1}\gamma_{2j}$; and (3) with probability $(1-p)(1-q)$ we do nothing, i.e. act with the identity gate.

{\bf Even time steps:} We perform 2-Majorana gates on all pairs $(2j,2j+1)$ $(j=1,\ldots,N)$ of nearest neighbor Majoranas.  For each such pair $(2j,2j+1)$ the gate is chosen randomly from $3$ possibilities: (1) with probability $p$ we act with $U_{2j,2j+1}$; (2) with probability $(1-p)(1-q)$ we measure $i\gamma_{2j}\gamma_{2j+1}$; and (3) with probability $(1-p)q$ we do nothing, i.e. act with the identity gate.

\begin{figure}
\centering
\includegraphics[width=0.8\linewidth]{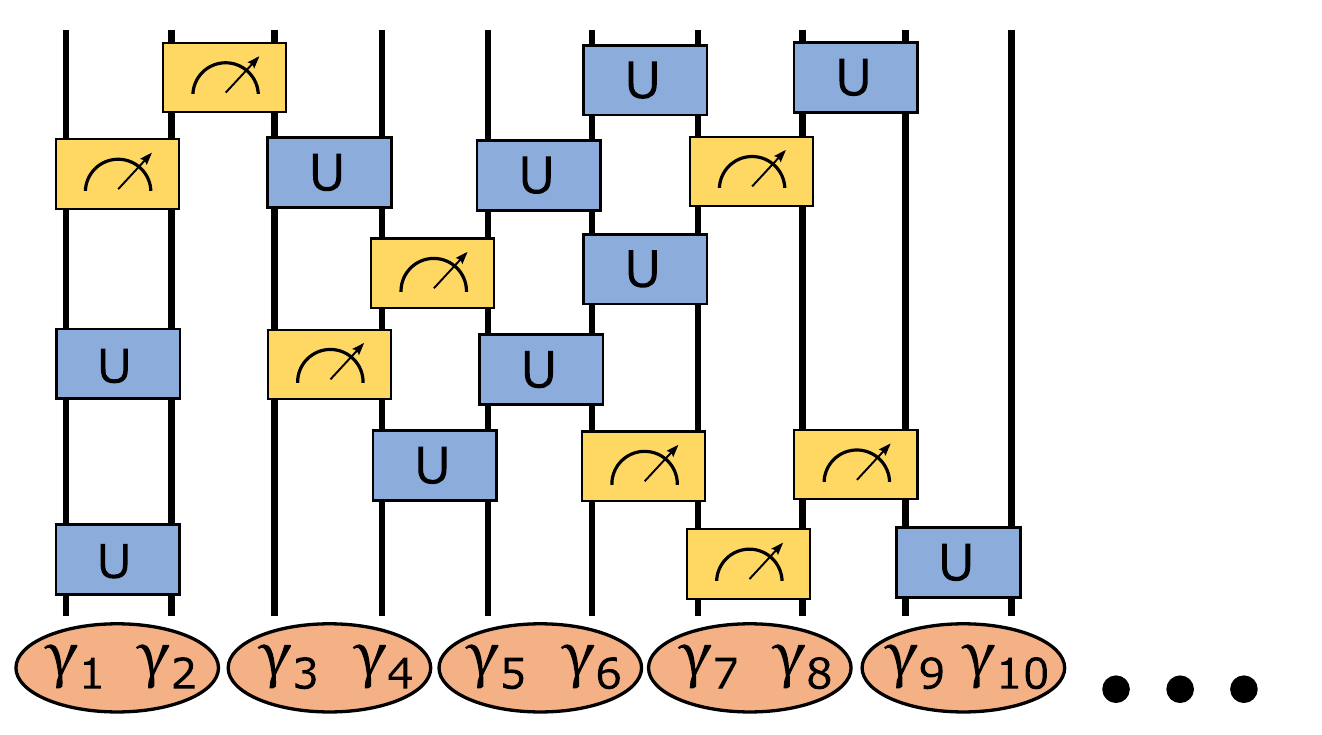}
\caption{(Color online) The protocol for the hybrid dynamics described in the text.  The vertical direction represents time; odd and even time steps correspond to the two complementary way of pairing up neighboring Majoranas; for each such pairing, nearest neighbor gates are applied which either perform a measurement (and record the outcome), apply a certain unitary gate, or do nothing.}
\label{fig:protocol}
\end{figure}

This protocol is illustrated in Figure \ref{fig:protocol}. 
%Because $q$ alternates with $(1-q)$ between time steps, we expect the model to have $q \leftrightarrow (1-q)$ symmetry in the thermodynamic limit.
We note here a symmetry: with periodic boundary conditions, sending $q\rightarrow (1-q)$ is equivalent to shifting each Majorana operator by one, $\gamma_k \rightarrow \gamma_{k+1}$; since none of the operations depend explicitly on the index $k$, this is a symmetry. Thus, the phase diagrams will have $q \leftrightarrow (1-q)$ symmetry.  We expect that with open boundary conditions, the symmetry will still hold in the thermodynamic limit.

\begin{figure}
    \centering
     \begin{minipage}{\linewidth}
    \includegraphics[width = 0.7\textwidth]{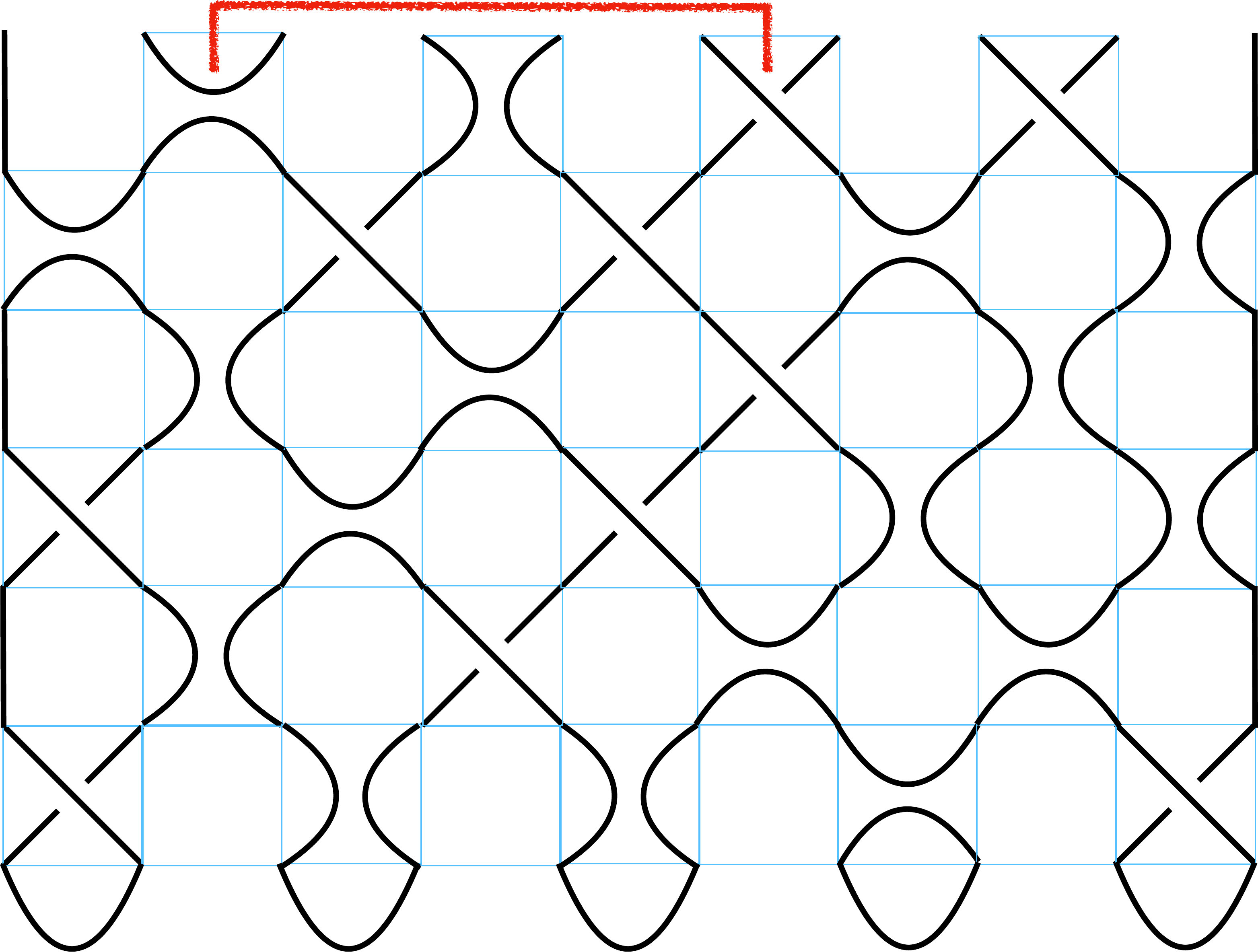}
    \caption{\label{fig:loops1}A trajectory of the circuit model with measurements, shown here with open boundary conditions.  Rotated by 45 degrees it becomes a configuration of the completely packed loop model with crossings (CPLC).  The black lines are Majorana worldlines.  The entanglement of the second and third physical fermion (red region) is $2 \log \sqrt{2} = \log 2$.}
    \end{minipage}
    \quad
    \begin{minipage}{\linewidth}
    \includegraphics[width = 0.7\textwidth]{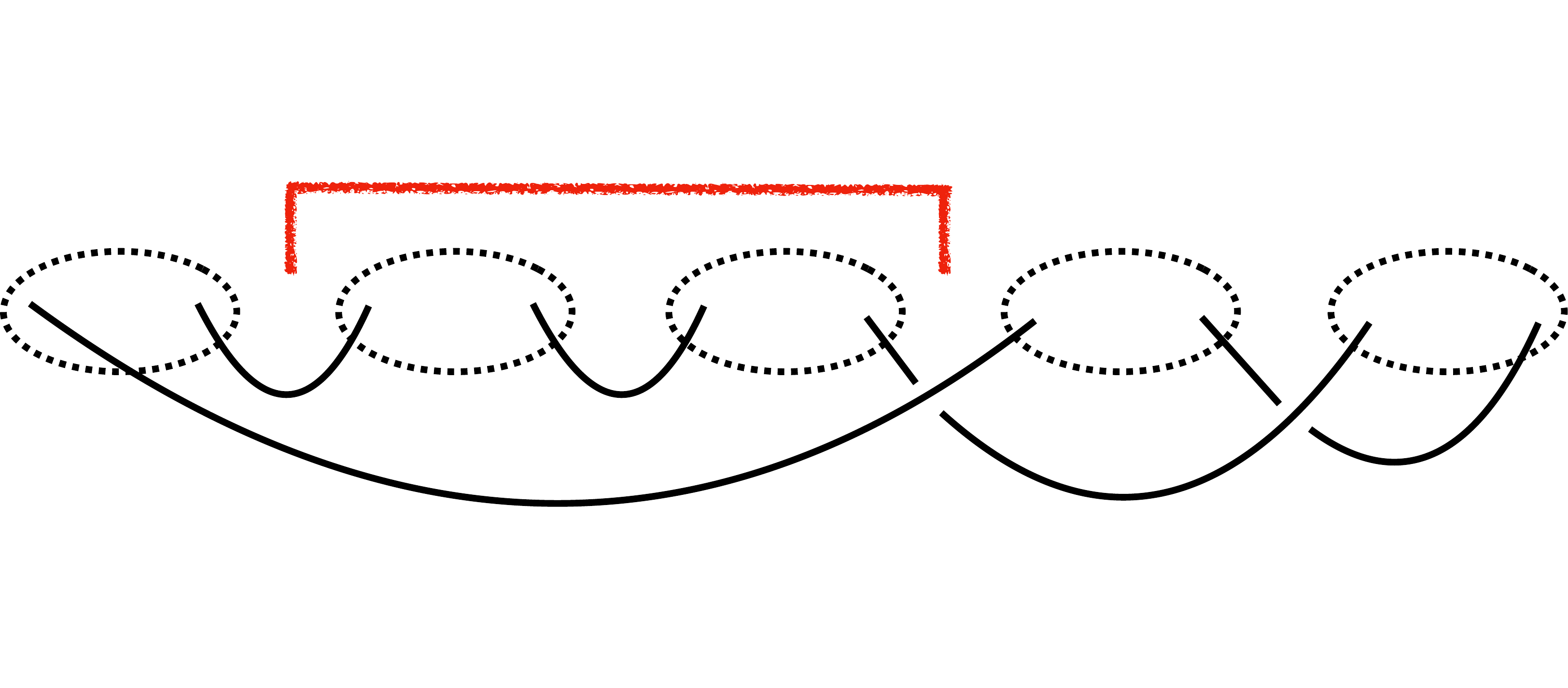}
    \caption{\label{fig:loops2}This simplified graphical view of the fermionic stabilizer state shows how the worldlines connect parts of the state in Fig. \ref{fig:loops1}.  The dotted ovals denote which Majoranas are paired up into physical fermions.}
    \end{minipage}
\end{figure}

\subsection*{Connection to Completely Packed Loop model with crossings}

Having defined our quantum model, we now describe how to map it exactly to a known statistical mechanical model, the completely packed loop model with crossings (CPLC).  To do this, it will be useful to introduce the notion of a fermionic stabilizer of a state $|\Psi\rangle$ in our $2^N$ dimensional many body Fock space. We define a set of stabilizers as a collection of $N$ commuting bilinears $i\gamma_{\sigma(2j-1)}\gamma_{\sigma(2j)}$, where $1\leq j \leq N$ and $\sigma$ is some permutation of $(1,\ldots,2N)$, such that $i\gamma_{\sigma(2j-1)}\gamma_{\sigma(2j)}|\Psi\rangle = \pm |\Psi\rangle$ for each $j$. A state is a stabilizer state if such a set of stabilizers exist.

The notion of a fermionic stabilizer state is useful because it is preserved by our dynamics: the quantum trajectory of an initial stabilizer state consists only of stabilizer states.  To see this, let us assume that we have a stabilizer state $|\Psi\rangle$, and let us act with one time step of our dynamics, analyzing the action of each of the gates in turn.  If the gate is the identity, certainly the stabilizer nature of the state is unchanged, since the state itself is unchanged.  If the gate is $U_{r,r+1}$, then, up to sign, this just exchanges the two Majoranas $\gamma_r$ and $\gamma_{r+1}$, so the state remains a stabilizer state, with stabilizer given by composing the permutation $\sigma$ by the exchange of $r$ and $r+1$.  Finally, let us analyze what happens when we measure $i\gamma_r \gamma_{r+1}$.  First, in the case when $r$ and $r+1$ are already paired by the stabilizer, the state is an eigenstate of $i\gamma_r \gamma_{r+1}$, so a measurement just reads off the eigenvalue but does not change the state.  Now suppose that $r$ and $r+1$ are not paired up by the stabilizer, so that $r$ is paired up with $s$ and $r+1$ is paired up with $t$.  For the purpose of measuring $i\gamma_r \gamma_{r+1}$, we may then imagine the system to just consist of $\gamma_r,\gamma_{r+1},\gamma_s, \gamma_t$, i.e. the system is the $4$-dimensional Fock space defined by these four Majorana operators.  This is because the remaining Majoranas are decoupled, with the state a (graded) tensor product between this $4$-dimensional system and that defined by the remaining Majoranas.  Now, since $|\Psi\rangle$ is an eigenstate of $\gamma_r \gamma_{r+1} \gamma_s \gamma_t$, measuring the commuting operator $i\gamma_r \gamma_{r+1}$ brings it to an eigenstate of both of these, and hence also of $i\gamma_s \gamma_t = -\left(i\gamma_r\gamma_{r+1}\right)\left(\gamma_r \gamma_{r+1} \gamma_s \gamma_t \right)$.  Hence the new state is stabilized by $i\gamma_r \gamma_{r+1}$ and $i\gamma_s \gamma_t$, together with all of the previous stabilizers. This can be seen as a type of entanglement swapping \cite{ref:swapping} between the Manorana fermions due to the joint measurement $i\gamma_r \gamma_{r+1}$. Note that this new set of stabilizers is independent of the measurement outcome\footnote{The Majorana content of the stabilizers, specifically. The measurement outcome affects an overall factor of $\pm 1$ on the stabilizers, which will be unimportant in our analysis.}.

The pairing rules so described can be tracked by representing the trajectory by a configuration of loops, as in Figure \ref{fig:loops1}.  The loops can be viewed as inhabiting a checkerboard pattern of squares, corresponding to an alternating sequence of even and odd steps in time (vertical direction).  Each square has one of three different types of configurations in it: an exchange (corresponding to a unitary gate), a configuration preserving the position of the two Majoranas (corresponding to doing nothing), and a `capped off' configuration (corresponding to measuring that Majorana bilinear).  Thus, by the discussion in the previous paragraph, the stabilizers of the final state are given by the pairing of the top endpoints in the diagram, as illustrated in Fig. \ref{fig:loops2}.

The probability of each configuration occurring is the product of the various probabilities $p$, $(1-p)q$ and $(1-p)(1-q)$ over all the squares of the checkerboard pattern.  This is just the Boltzmann weight of the completely packed loop model with crossings (CPLC) \cite{Nahum}.   

\subsection*{Spanning Number and Entropy}

In the previous subsection we outlined a correspondence between a model of free fermion hybrid dynamics and the CPLC.  In order for this correspondence to be useful, we have to identify corresponding observables on the two sides.  On the CPLC side, the observable we will compute is the spanning number $n_s$ on a cylinder, with periodic boundary conditions in space, open boundary conditions in time, and both space and time having length $L$.  The spanning number is defined as the number of Majorana worldlines that connect the two boundaries of the cylinder (i.e. `span' the cylinder).  Its average is computed by weighing the various CPLC configurations with the Boltzmann weights defined above.

The spanning number has an appealing physical interpretation in our free fermion model in terms of entropy. For specificity we will work with the second Renyi entropy, but we note that for fermionic stabilizer states the Renyi and von Neumann entropies all coincide. The clearest way to formulate this is by bringing in another `ancilla' copy of the many-body Hilbert space, spanned by ancilla Majorana fermions $\gamma_j'$ ($1\leq j \leq 2N)$, and taking an initial state $|\Psi_0\rangle$ which is stabilized by $i\gamma_j\gamma_j'$ ($1\leq j \leq 2N)$.  The portion of this state on the original Hilbert space is then maximally mixed, and the free fermion dynamics may purify it to some extent, since it includes measurements.  At the final time $L$, the entanglement entropy between the system and the ancilla is given by $\frac{1}{2}\log 2$ times the number of stabilizers of the final state that link the system and ancilla. The number of such stabilizers is simply the number of worldlines that link the bottom (initial) and top (final) edge of the spacetime, i.e. just the spanning number $n_s$.  Thus the final state entropy, averaged over quantum trajectories, is precisely the average spanning number, up to the factor of $\tfrac{1}{2}\log 2$.

\begin{figure}
\centering
\begin{minipage}{0.99\linewidth}
\includegraphics[width=8.4cm]{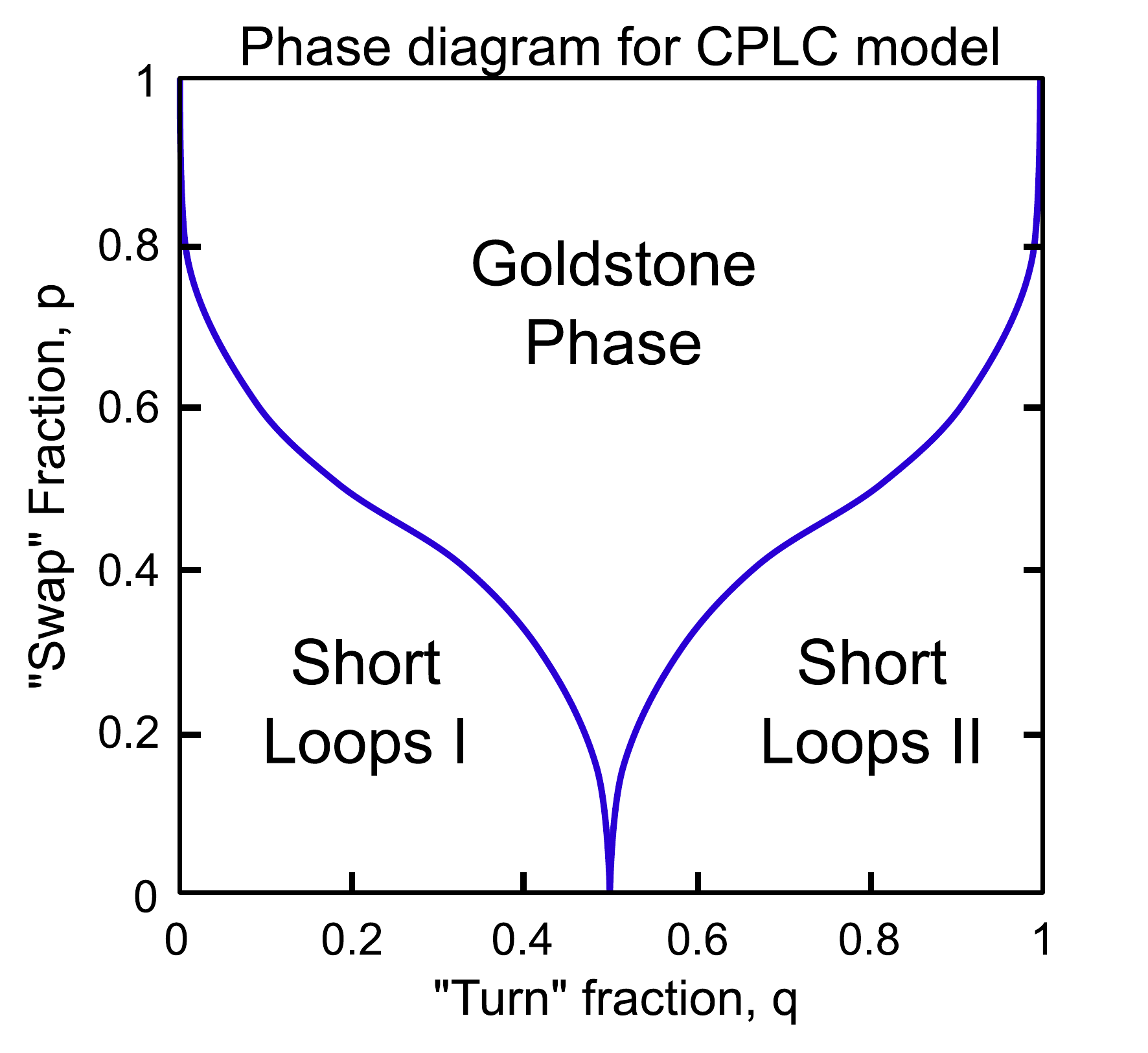}
\caption{(Color online) Phase diagram for the CPLC model \cite{Nahum}. The behavior of the average spanning number, $n_s$, determines the phase. In the short loops phase, $\langle n_s \rangle$ decreases with increasing system size. In the Goldstone phase, $\langle n_s \rangle$ increases logarithmically with system size. The spanning number generalizes to the second R\'enyi entropy in our more general fermion model.}
\label{fig:CPLC}
\end{minipage}

\vspace{5mm}

\begin{minipage}{0.99\linewidth}
\includegraphics[width=8.4cm]{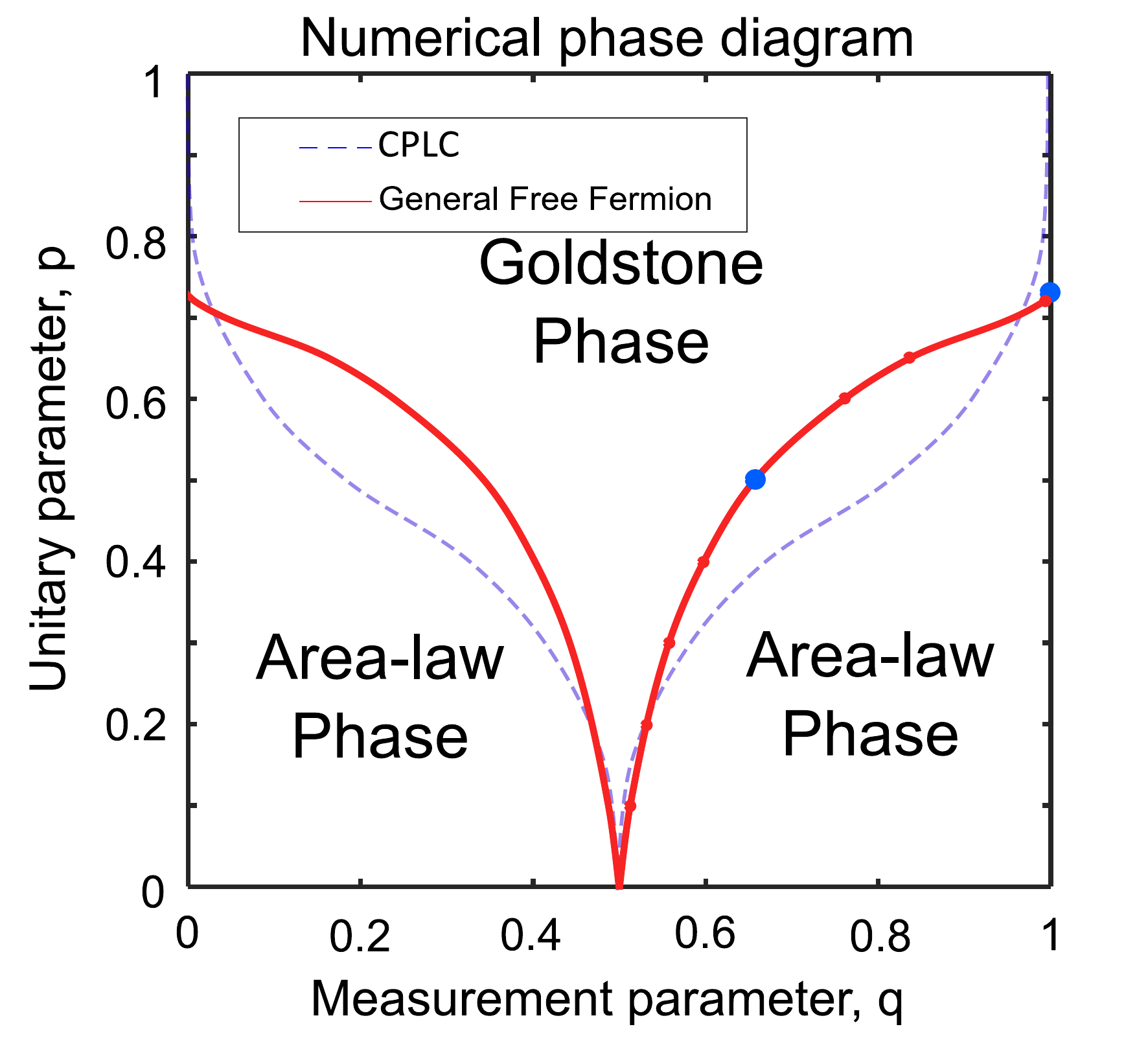}
\caption{(Color Online) Phase diagram for more general free fermion model, with CPLC phase diagram for reference. The large blue dots are locations in the phase diagram which are studied in more depth in this paper. Note that on the top line, $p = 1$, all points represent the same phase point, namely, the volume-law phase consisting of purely unitary evolution. Note the ${q \leftrightarrow (1-q)}$ symmetry, mentioned in Section \ref{sec:model}.}
\label{fig:ffphase}
\end{minipage}

\end{figure}

Reference \cite{Nahum} shows that the spanning number is an order parameter for the phases that appear in the CPLC.  Specifically, there are two phases, as shown in Figure \ref{fig:CPLC}: the short loop phase and the Goldstone phase, with the spanning number scaling like $0$ and $\log L$ respectively, in the thermodynamic limit defined above.  Hence, in the hybrid dynamics the short loop phase corresponds to the area law phase.  Reference \cite{Nahum} explores several features of this phase diagram.  In particular, it is noted that at $p=0$ (i.e. the case of measurements only) the CPLC reduces to a model of the bond percolation transition tuned by $q$.  Furthermore, at both $q=0$ and $q=1$, it is shown that the field theory describing the CPLC model possesses an extra $U(1)$ symmetry which guarantees that the short loop phase extends all the way to $p=1$.  At values of $q$ different from $0,\frac{1}{2},1$, there is a transition from the short loop to the Goldstone phase at some $p$ with $0<p<1$.  Reference \cite{Nahum} studies this phase trasition at $p= \frac{1}{2},q=0.82$ and extracts a correlation length scaling exponent $\nu=2.745(19)$.

The goal for the remainder of this work is to examine the extent to which the CPLC phase diagram, viewed in terms of the hybrid dynamics, is robust to more general free fermion dynamics.  In order to pursue this, we now introduce the formalism of Gaussian states.

\section{More general free fermion models - Gaussian state formalism} \label{sec:gaussian}

To generalize beyond the specific case of the CPLC dual, we will replace the unitary gate $U_{r,r+1}$, which, up to sign, just swaps $\gamma_r$ and $\gamma_{r+1}$, with a more general local free fermion unitary gate.  Namely, we will consider unitary operators of the form $U=\exp\left(\frac{1}{4} A_{ij} \gamma_i \gamma_j\right)$ with $A$ a real anti-symmetric $2n$ by $2n$ matrix.  The action of $U$ by conjugation is an orthogonal rotation of the $2n$ Majoranas:

\begin{align*}
    U \gamma_i U^{-1}=\exp(-A_{ij}) \gamma_j
\end{align*}
As far as measurements, the most general general measurements we might want to consider are of the fermion linear optics (FLO) type \cite{Bravyi}, which can always be thought of as projecting onto eigenstates of some $i\gamma_a \gamma_b$ operator, after appropriate basis transformation.  However, for this work we will simply retain the exact same measurements as in the CPLC dual model introduced above.

The generalization we will investigate involves taking the same protocol as above, but for the unitary gates, instead of applying the fixed unitary $U_{r,r+1}$ defined above, we draw one randomly from the class of all two-Majorana unitaries.  All such unitaries have the action:

\begin{align} \label{eq:rotate}
    \gamma_1 &\rightarrow \cos(\alpha) \gamma_1 - \sin(\alpha)\gamma_2 \\ \nonumber
    \gamma_2 &\rightarrow \sin(\alpha) \gamma_1 + \cos(\alpha)\gamma_2
\end{align}
for some $\alpha$; the case of the $U_{r,r+1}$ defined above corresponds to $\alpha= -\frac{\pi}{2}$.  In our general model we will draw $\alpha$ randomly from between $0$ and $2 \pi$.

Because acting with a generic unitary gate of the above form now maps a Majorana operator to a superposition of Majorana operators, we can no longer apply the stabilizer formalism of Section \ref{sec:model} to relate entanglement entropy to spanning number.  Instead, we will perform numerical Monte Carlo studies of the hybrid dynamics.  The numerics will start with a maximally mixed state $\rho_0$ and apply

\begin{align}
    \rho_0 \rightarrow C \rho C^\dagger / \rm{Tr}\, C \rho_0 C^\dagger
\end{align}
where $C$ is a (non-unitary) circuit made up of a product of randomly chosen unitaries and projectors corresponding to a particular quantum trajectory.  Specifically, the probability density associated to a particular circuit $C$ is equal to the CPLC Boltzmann weight (a product of powers of $p,q,(1-p)$ and $(1-q)$), multiplied by the uniform probability density associated with the choice of each unitary, multiplied by the Born probability $\rm{Tr}\, C \rho_0 C^\dagger$ associated to the measurement outcomes in $C$.  Even though the states appearing in the quantum trajectories are no longer stabilizer, they still have the property of being `Gaussian', and this allows for efficient numerical simulation.  We now review the formalism of these Gaussian states.

\subsection*{Gaussian state formalism}

This exposition follows reference \cite{Bravyi} closely.  First, let us define a Gaussian state.  A mixed state $\rho$ can be viewed as an operator, and, as such, has some expansion in polynomials in the $\gamma_i$.  Given such an expansion, with each $\gamma_i$ appearing to a power $0$ or $1$ in each term, we can form an associated element of a Grassmann algebra by replacing each $\gamma_i$ with a Grassmann number $\theta_i$.  $\rho$ is then called Gaussian if the corresponding Grassmann algebra element is of the form

\begin{align*}
    \frac{1}{2^N}\exp\left(\frac{i}{2} \theta^T M \theta\right)
\end{align*}
where $M$ is a $2N$ by $2N$ real anti-symmetric matrix, called the covariance matrix of the state.
%Each such $M$ can be transformed, by an orthogonal rotation, to block diagonal form with two by two anti-symmetric blocks with $\pm \lambda_i$ on the diagonal, where $-1\leq \lambda_i \leq 1$ for all $i$.  The $|\lambda_i|$ are called the Williamson eigenvalues of $M$, and a pure state corresponds to all $|\lambda_i|=1$.
Each such $M$ can be transformed, by an orthogonal rotation, to block diagonal form with $N$ two-by-two blocks on the diagonal. Each block is anti-symmetric, and so determined by a number $\lambda_i$ on the off-diagonal, where ${-1\leq \lambda_i \leq 1}$ for all $i$. The $|\lambda_i|$ are called the Williamson eigenvalues of $M$, and a pure state corresponds to all $|\lambda_i|=1$.

Let us see how a Gaussian state evolves under our hybrid evolution.  First, evolving $\rho$ under free fermion unitary transformations is easy: we just conjugate $M$ by the rotation in Eq. (\ref{eq:rotate}) (rotating on the appropriate $2d$ subspace of the $2N$ Majoranas, and acting as the identity on the complement).  The result is the covariance matrix of the new state, which remains Gaussian.  Now let us consider evolving $\rho$ under a measurement of $i \gamma_j \gamma_{j+1}$ (with post-selection, i.e. projecting onto an eigenspace of $i \gamma_j \gamma_{j+1}$ and normalizing).  Reference \cite{Bravyi} shows that in this case the normalized, post-selected post-measurement state remains Gaussian.  Its covariance matrix $M'$, for the measurement outcome $s=\pm 1$, is determined as follows.  Let $K$ be the anti-symmetric matrix whose entries $(p,q)$ (i.e. row $p$ column $q$) are $(\delta_{p,j} \delta_{q,j+1} - \delta_{p,j+1}\delta_{q,j})$.  Let $L=(I-sMK)^{-1} M$.  Then the $M'_{p,q} = L_{p,q}$ if $p,q \notin \{j,j+1\}$, and $M'_{p,q} = sK_{p,q}$ otherwise.  This turns out to have all Williamson eigenvalues equal to $+1$ if $M$ does as well, so pure states indeed evolve into pure states.  The probability of the outcome $s$ is $\frac{1}{2} \rm{Pf} (M)\, \rm{Pf} \,(sK-M^{-1})$, where $\rm{Pf}$ denotes the Pfaffian.

We can try to simplify the equation for $L$ using the Taylor expansion. Since $K$ is only nonzero in the ${j,j+1}$ block, we find that $KMK = -M_{j,j+1}K$. Thus, $(MK)^n = (-M_{j,j+1})^{n-1} MK$, and a Taylor expansion of the form $(1-x)^{-1}$ gives
\begin{comment}
\begin{align}
    \nonumber (1 - sMK)^{-1} &= 1+\sum_{n=1}^\infty(MK)^n \\
    %\nonumber &= 1 + \sum_{n=0}^\infty \left( s^n M(-M_{j,j+1})^{n-1} K\right)\\
    \nonumber &= 1 + sMK\sum_{n=0}^\infty (-sM_{j,j+1})^n\\
    \label{eq:inv}&= 1+\frac{sMK}{1+sM_{j,j+1}}
\end{align}
\end{comment}
\begin{equation}
    \label{eq:inv}
    (1 - sMK)^{-1} = 1+\frac{sMK}{1+sM_{j,j+1}}
\end{equation}
There are limits on when such a Taylor expansion is justified; however, it can be shown by direct substitution that (\ref{eq:inv}) is indeed the inverse of $(1-sMK)$. Thus,
\begin{equation*}
    L = M + \frac{sMKM}{1+sM_{j,j+1}}
\end{equation*}
Note that performing the evolution takes resources which are polynomial in $N$, since we just have to follow the covariance matrix rather than the full quantum many body state.  This is the advantage of the Gaussian state formalism.

In terms of the Williamson eigenvalues $\lambda_i$, $i=1,\ldots, N$ of $M$, the $2^N$ many body Schmidt eigenvalues are

\begin{align}
    \frac{1}{2^N} \prod_{i=1}^N \left(1 \pm \lambda_i \right)
\end{align}
where each eigenvalue corresponds to one of the $2^N$ choices of sign for the $\pm \lambda_i$. The von Neumann entanglement entropy of the many body state is thus
\begin{align}
    S_2 &=  - \sum_i \left[\frac{1-\lambda_i}{2} \log \left(\frac{1-\lambda_i}{2}\right) + \frac{1+\lambda_i}{2} \log \left(\frac{1+\lambda_i}{2}\right)\right]
\end{align}
which can be extracted by diagonalizing the covariance matrix.  The second Renyi entropy is

\begin{align*}
    S_2 &= -\tfrac{1}{2} \log \text{Tr} \rho^2\\
    &= N \log 2 - \sum_i \log (1+\lambda_i^2) \\
    &= N \log 2 - \tfrac{1}{2}{\rm{Tr}}\, \log \left(1-M^2 \right)
\end{align*}
which can be extracted directly from the covariance matrix.

\section{Results} \label{sec:results}

\subsection*{Phase diagram}

\begin{figure}[t]
	\centering
	\begin{minipage}{8.6cm}
		\includegraphics[width = \textwidth]{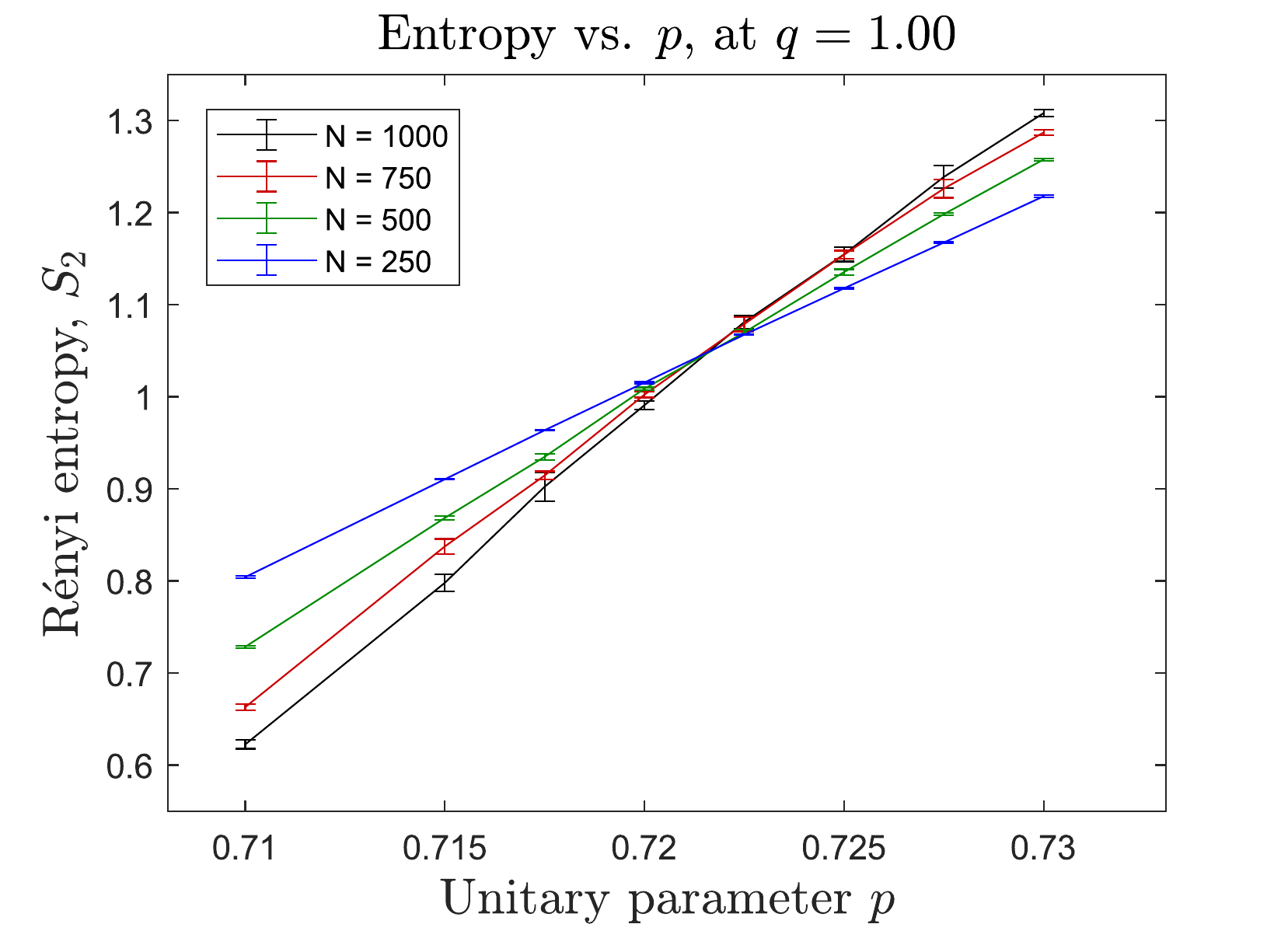}
	\end{minipage}
	
	\vspace{5mm}
	
	\begin{minipage}{8.6cm}
		\includegraphics[width = \textwidth]{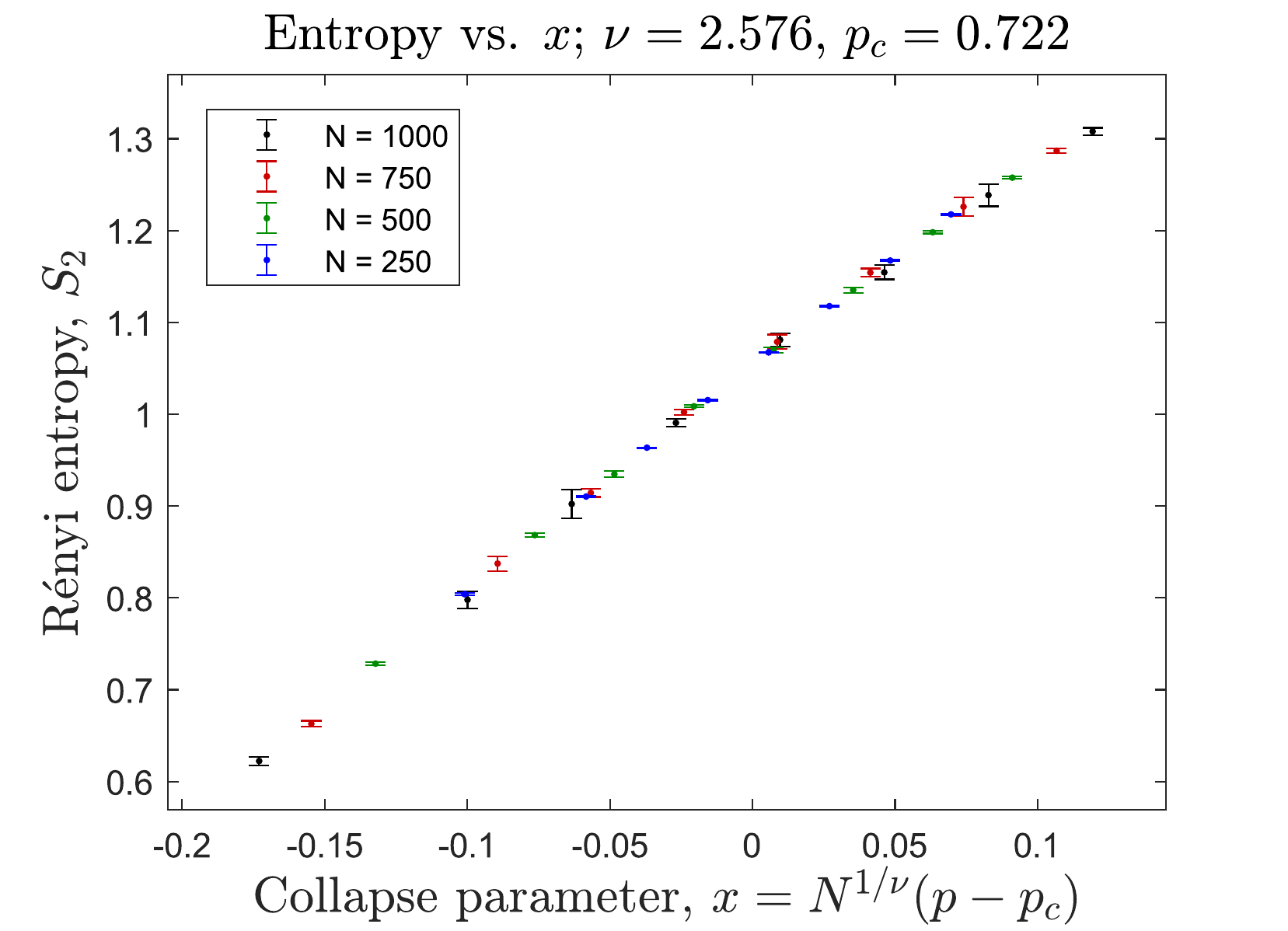}
	\end{minipage}
	\caption{\label{fig:q1Trns}(Color Online) Graphs of $S$ for $q = 1.00$, $p$ near $0.72$ for the generalized free fermion model. The data collapse values for $p_c$ and $\nu$ were calculated by fitting the data to a fifth-order polynomial for various values of $p_c$, $\nu$ and finding the values which minimized the residual sum of squares (sum of squares error). The data collapse indicates the presence of a continuous phase transition for $q=1.00$ near $p = 0.72$, a feature of the generalized free fermion model that is not shared with the CPLC.}
\end{figure}

\begin{figure}[t]
	\centering
	\includegraphics[width=8.4cm]{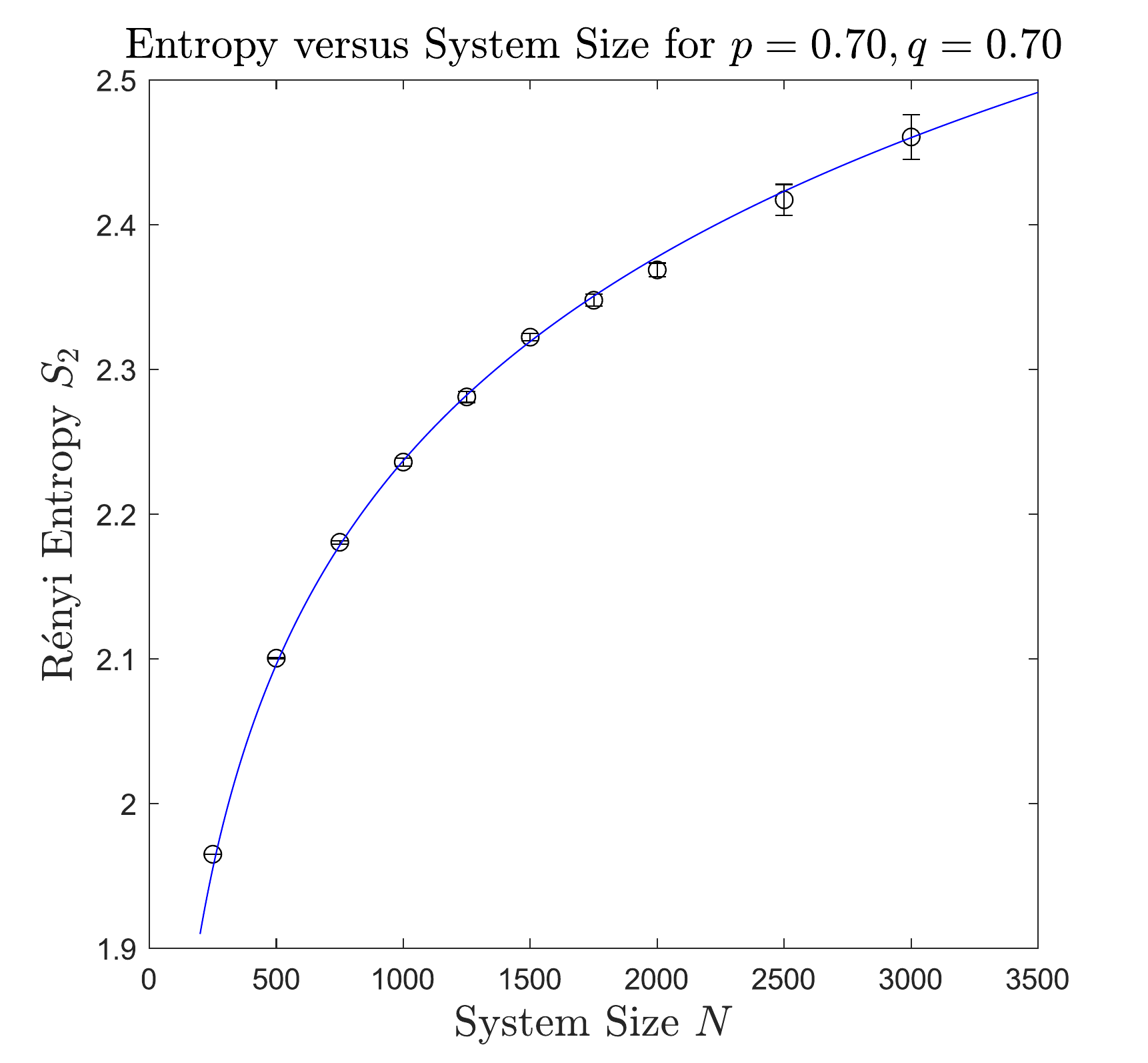}
	\caption{Entropy law for a point inside the Goldstone phase in the generalized free fermion model. The blue line is a fit to the logarithm $a\log(bx)+c$. The values obtained were: $a=0.203$, $b=0.281$, $c=1.091$. The 95\% confidence bounds on $a$ are $(0.1957, 0.2107)$.}
	\label{fig:GLD7}
\end{figure}

The model was numerically simulated on the Hyak supercomputer at the University of Washington. For each system size, the evolution of 40 independent systems was run in parallel using Matlab's Parallel Computing Toolbox.

Our main results are summarized in the phase diagram given in Fig. \ref{fig:ffphase}. We find that the two distinct phases persist in the more general free fermion model, though the line separating them has changed. Fig. \ref{fig:GLD7} shows the entropy scaling for the point $p=0.7,q=0.7$, inside the Goldstone phase, demonstrating that the logarithmic-law scaling persists in the more general model.

\begin{figure}
	\centering
	\includegraphics[width=8.6cm]{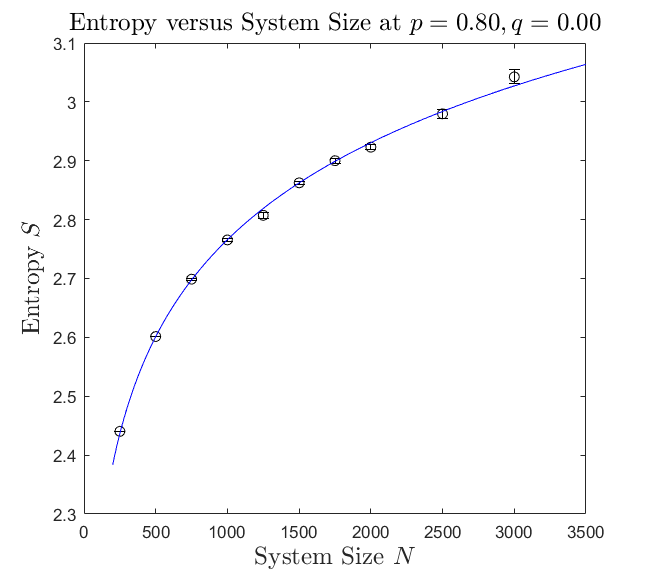}
	\caption{Entropy law for a point on the edge of the phase diagram ($q=0.00$). This point is in the area-law phase (short loops phase) of the CPLC model, but inside the Goldstone phase for the general model. The blue line is a fit to the logarithm $a\log(bx)+c$. The values obtained were: $a=0.238$, $b=0.646$, $c=1.228$. The 95\% confidence bounds on $a$ are $(0.2282, 0.2473)$.}
	\label{fig:GLD8}
\end{figure}

One of the distinct differences between the models is the existence of a phase transition on the boundaries of the phase diagram, when $q = 0,1$. In the CPLC model, these values of $q$ introduce an extra symmetry on the model, preventing a phase transition \cite{Nahum}. However, we find a phase transition at $p = 0.72(2)$, $q = 0,1$, shown in Fig. \ref{fig:q1Trns}. The entropy scaling for the point $p=0.8$, $q=0$ is shown in Fig. \ref{fig:GLD8}, a point which would be in the area-law phase of the CPLC model. We instead see a logarithmic-law scaling, demonstrating that it is instead in the Goldstone phase of our model.

It's also worth noting that these new phase transitions occur at the most experimentally feasible parts of the phase diagram. At $q=0$, all measurements in the evolution are of on-site fermion parity, with no inter-site projective measurements. Further, the measurements all mutually commute, giving an entanglement transition that uses commuting projective measurements.

\subsection*{Extracting $\nu$ at the generic transition}

A finite-size data collapse helps to confirm that this model shows properties of critical phenomena. The underlying idea is that near a phase transition, thermodynamic properties should scale as a universal function of $L/\xi$, where $L$ is the (linear) system size and $\xi$ is the correlation length \cite{Beach, FisherFSS}. For large system size and small reduced temperature $t = (T-T_c)/T_c$ around the critical temperature $T_c$, a thermodynamic observable $Q$ should go as
\begin{equation*}
    Q(L,T) = L^{\kappa/\nu} f(tL^{1/\nu})
\end{equation*}
for some function $f$. The intuition for this one parameter scaling form is that as we approach the scale-invariant critical point, a change in length scale can be compensated by a change in temperature. Here, $\nu$ is the correlation length exponent.

In our work, the tuning parameter $T$ is not the temperature but rather $p$ or $q$, depending on context. Thus, we define
\begin{equation*}
    x = N^{1/\nu}(T-T_c)
\end{equation*}
and make plots of the entropy $S$ versus this parameter $x$. By varying the values of $\nu$ and $T_c$, we attempt to find the value that gives the best collapse of the data points onto a single line in the plot.

\begin{figure}[t]
	\centering
	\begin{minipage}{8.6cm}
		\includegraphics[width = \textwidth]{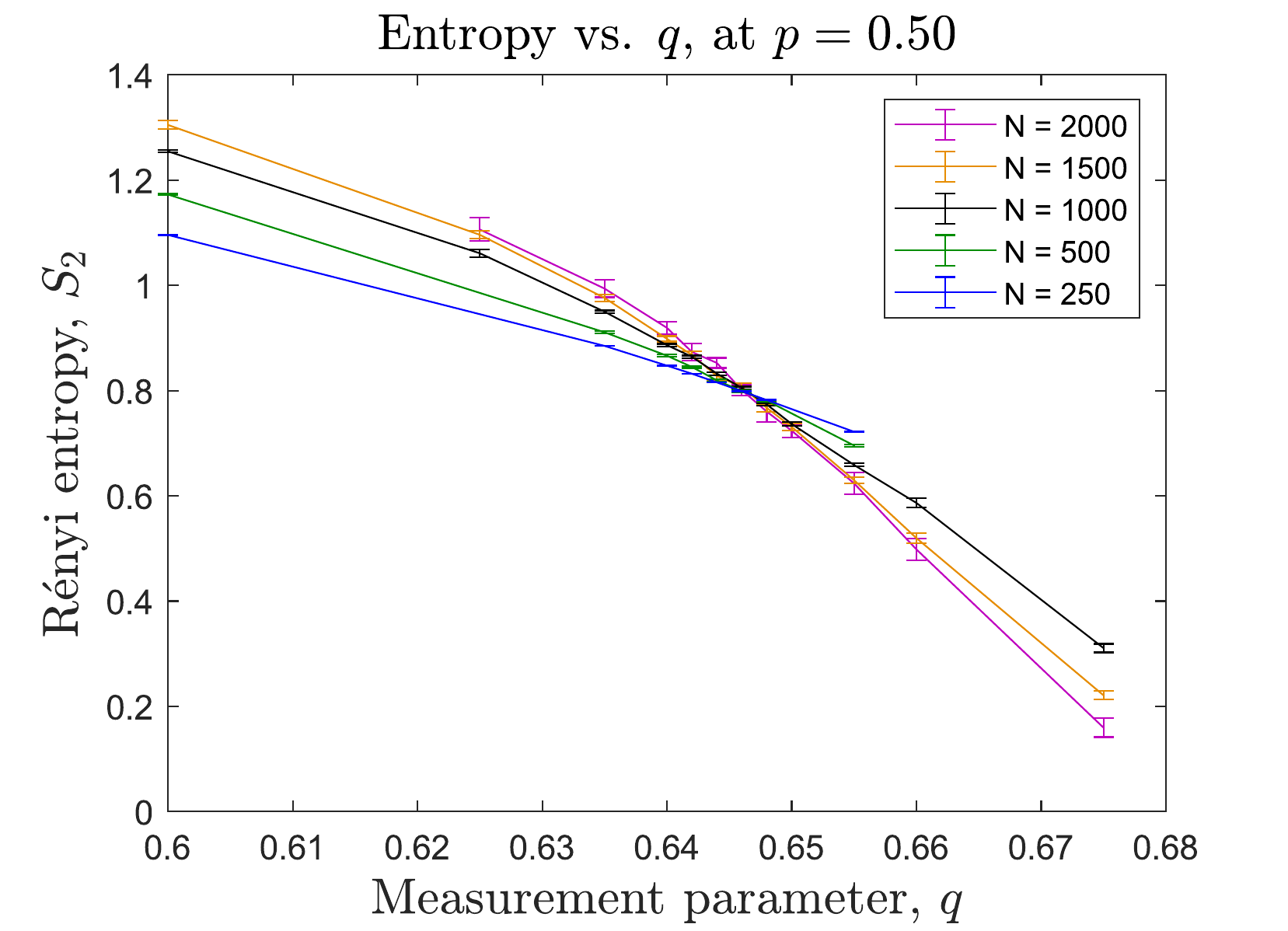}
	\end{minipage}
	
	\vspace{5mm}
	
	\begin{minipage}{8.4cm}
		\includegraphics[width = \textwidth]{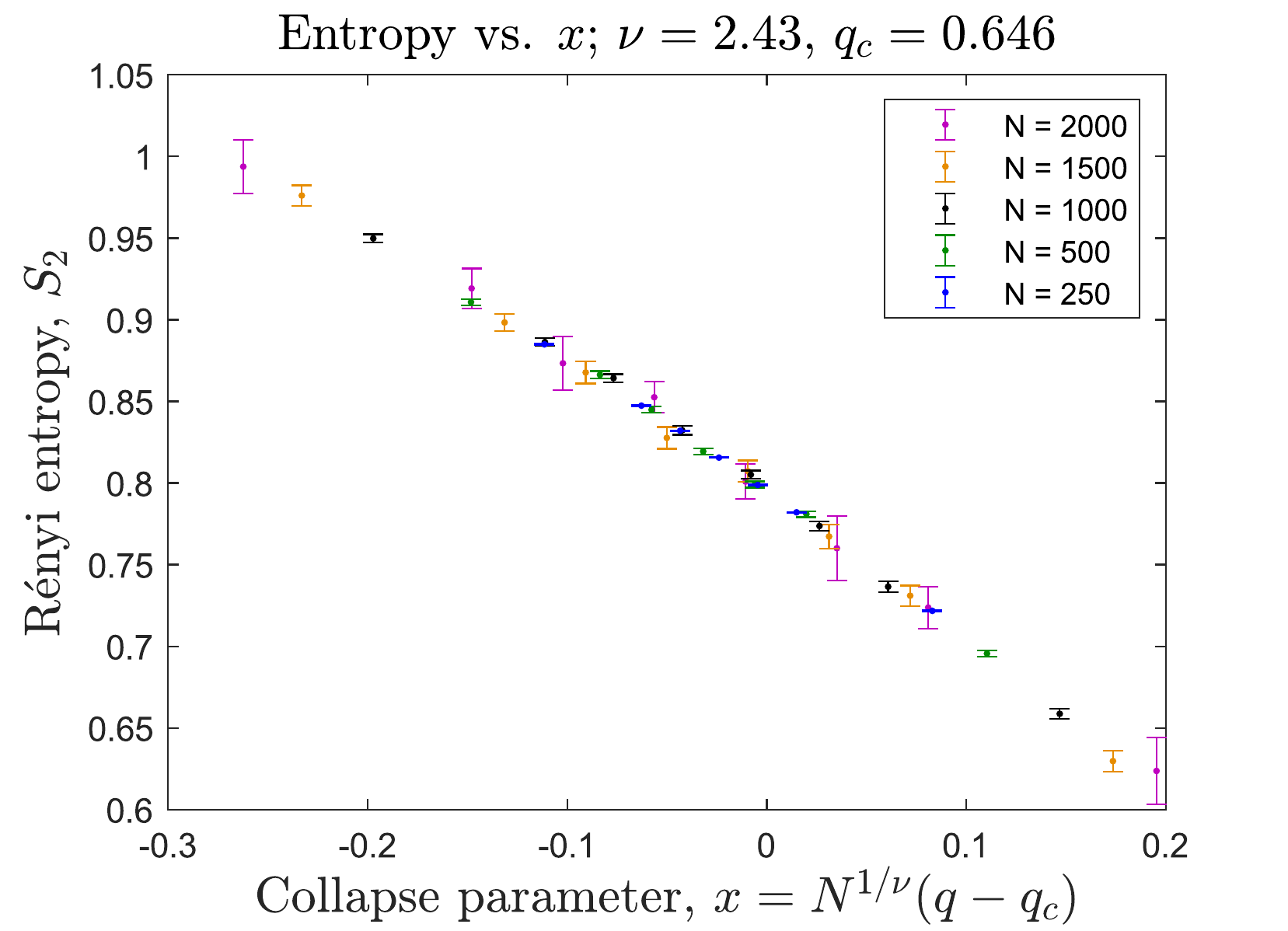}
	\end{minipage}
	\caption{\label{fig:maingraph}(Color Online) Graphs of $S$ for $p = 0.50$, $q$ near $0.64$ in the general free fermion model. The data collapse values for $q_c$ and $\nu$ were calculated by fitting the data to a fifth-order polynomial for various values of $q_c$, $\nu$ and finding the values which minimized the residual sum of squares (sum of squares error).}
\end{figure}

We first study the case $p=0.50$, with $q$ playing the role of $T$.  The results for this case are shown in Fig. \ref{fig:maingraph}. We get the values for $\nu$ and $q_c$ by minimizing the error in fitting the collapsed data to a fifth order polynomial. We obtain values
\begin{align}
    q_c &= 0.64(6) & \nu &\approx 2.41(6)
\end{align}
We note that the value for $q_c$ is smaller than for the $p=0.5$ transition in the CPLC model, which is at $q_{c,CPLC} \approx 0.82$. The value of $\nu$ is also smaller, with $\nu_{CPLC} \approx 2.745$ at $p=0.5$ \cite{Nahum}. However, the uncertainties in $\nu$ from numerical errors and possible irrelevant variables leave us unable to definitively rule out the possibility that our model is in the same universality class as the CPLC transition.

Further, we find a phase transition for the boundaries of the phase diagram, $q = 0,1$, where none exists in the CPLC model. The data for this is shown in Fig. \ref{fig:q1Trns}. Here, we obtain values
\begin{align*}
    p_c &= 0.72(2) & \nu &\approx 2.5(8)
\end{align*}
This $\nu$ is larger than the value found at $p = 0.5$, though still less than what was found for the CPLC model. Again, this is consistent with the points being in the same universality class.

\section{Discussion} \label{sec:discussion}

In this work we have investigated the purification transition in certain fermionic models of hybrid unitary-measurement dynamics in one dimension.  We used an exact duality from the known CPLC statistical mechanical model to understand one particular case, and then numerically investigated a particular generalization away from this tractable point.
%[\textbf{\textit{What does ``exactly solvable'' mean in this context?}}]
We found that the generalized model retains all of the phases present in the exactly solved model, albeit with slightly shifted phase boundaries.  Although we only investigated one specific generalization away from the exactly solvable fixed point, we expect this robustness to persist in general.

One outstanding question that remains is to find a statistical mechanical dual for a general free fermion model in the above class.  If one is interested specifically in say the second Renyi entropy, one may relatively easily write an expression for it as a certain replica limit of a ratio of quantities which have the interpretation of matrix elements of imaginary time evolution operators, following Section VI of \cite{NahumLoops}.  Following the strategy used in the usual entanglement transition \cite{Vasseur} one may then hope to interpret these quantities as partition functions of statistical mechanical model.  One way to do this is to insert resolutions of the identity on replicated sites which are roughly integrals over $SO(2N)$ of $SO(2N)$-rotated projectors.  Although it will certainly be difficult to make rigorous statements about the replica limit, one may hope that at least some symmetry based arguments can be made.  For example, one may hope to explain the existence of a phase transition at $q=0,1$ in the generalized models, in contrast to the lack of such a transition in the CPLC dual model, by showing that the CPLC dual has an enhanced symmetry at $q=0,1$ (see also \cite{Nahum}).

In addition, the CPLC model distinguishes between the two short-loops phases, see Fig. \ref{fig:CPLC}. The ``Short Loops II'' phase at ${q > \tfrac{1}{2}}$ is distinguished by the existence of a macroscopic loop which circles the configuration when open boundary conditions are used. This is tied to a topological phase in \cite{Nahum}. The two area-law phases in the generalized model are also tied to the same topological phases, which in the quantum system can be distinguished e.g. by measuring a string order parameter.  How to define and numerically measure such an order parameter in an associated statistical mechanical model is a question that we leave for future investigation.

\section*{Acknowledgements} 
This work was facilitated through the use of advanced computational, storage, and networking infrastructure provided by the Hyak supercomputer system and funded by the Student Technology Fee at the University of Washington. JM and LF were also supported by NSF DMR-1939864.

\bibliography{References} % bib only shows references actually used in the paper.

\end{document}